\documentclass{jltp}
\usepackage{amssymb}
\usepackage{graphicx}

\runninghead{V.~S.~L'vov, S.~V.~Nazarenko and L.~Skrbek}{Energy
Spectra of Developed Turbulence
 in Helium Superfluids}

\usepackage{amsmath,bm,epsfig}

\def\Fbox#1{\vskip1ex\hbox to 8.5cm{\hfil\fboxsep0.3zcm\fbox{%
  \parbox{8.0cm}{#1}}\hfil}\vskip1ex\noindent}  

\let \nn  \nonumber
\newcommand{\br}{\\ \nn}

\let\*\cdot
\def\<{\left\langle} \def\>{\right\rangle} \def\({\left(}
\def\){\right)}
\let\p\partial \let\~\widetilde \let\^\widehat 
  
  \def\1{\bm1} 
 
\newcommand{\B}[1]{{\bm{#1}}}
\def\BE{\begin{equation}}\def\EE{\end{equation}}
\def\BEA{\begin{eqnarray}}\def\EEA{\end{eqnarray}}
\def\BSE{\begin{subequations}}\def\ESE{\end{subequations}}

\renewcommand{\sb}[1]{_{\text {#1}}}  
\newcommand{\Sb}[1]{_{_{\text {#1}}}} 
  \def\Sb#1{_{\scriptscriptstyle\rm{#1}}}

\newcommand{\Ref}[1]{(\ref{#1})}
\newcommand{\REF}[1]{Eq.~(\ref{#1})}

\renewcommand{\a}{\alpha}
\renewcommand{\d}{\delta}
\newcommand{\D}{\Delta}\newcommand{\ve}{\varepsilon}
\renewcommand{\o}{\omega}

\let\p\partial

\let \= \equiv

\let\*\cdot
 \def\sb#1{_{\rm{#1}}}

 \def\({\left(} \def\){\right)}
 \def \[ {\left [} \def \] {\right ]}
   
    \let\^\widehat
  \let\-\overline

  \def\REF#1{Eq.~\Ref{#1}}
   \def\Ref#1{(\ref{#1})}  \def\<{\left\langle}
   \def\>{\right\rangle}

   \let\~\widetilde

\usepackage{graphicx}

\usepackage{rotating}

\begin {document}
\title{Energy Spectra of Developed  Turbulence
 in Helium Superfluids}
\author{V.S.~L'vov$^*$, S. V. Nazarenko$^\dag$ and
L.~Skrbek$^{\ddag,\S}$,   }
\address{$^*$ Department  of Chemical Physics, The Weizmann
Institute of
Science,\\ Rehovot 76100, Israel\\
$^\dag$   University of
Warwick, Mathematics Institute, Coventry, CV4 7AL, UK\\
$^\ddag$ Institute of Physics ASCR, Na Slovance 2, 182 21 Prague,
 Czech Republic\\
 $^\S$ Faculty of Mathematics and Physics,
 Charles University,\\ V  Hole\v{s}ovi\v{c}k\'ach 2,
 180 00 Prague, Czech Republic  }
\date{\today}
\maketitle
\begin {abstract}
We suggest a  ``minimal model"  for the 3D turbulent energy spectra
 in superfluids, based on their  two-fluid description.  We start from
the Navier-Stokes equation for the normal fluid and from the
coarse-grained hydrodynamic equation for the superfluid component
(obtained from the Euler equation for the superfluid velocity
after averaging over the vortex lines) and introduce a mutual
friction coupling term, proportional to the counterflow velocity,
the average superfluid vorticity and to the temperature dependent
parameter $q=\alpha/(1+\alpha')$, where $\alpha$ and $\alpha'$
denote the dimensionless parameters characterizing the mutual
friction between quantized vortices and the normal component of
the liquid. We then derive the energy balance equations, taking
into account the cross-velocity correlations. We obtain all
asymptotical solutions for normal and superfluid energy spectra
for limiting cases of small/big normal to superfluid density ratio
and coupling. We discuss the applicability of our model to
superfluid He II and to $^3$He-B.

\end{abstract}
\maketitle

 \section*{ INTRODUCTION}
Quantum turbulence\cite{VinenNiemela} containing a tangle of
singly quantized vortex line -- such as turbulence in superfluid
$^4$He (He-II) and in the superfluid B-phase of $^3$He ($^3$He-B)
-- besides of being for half a century\cite{VinenOld} a playground
for low temperature physicists increasingly attracts attention of
the fluid dynamics community. In this paper, we address the
important question of the turbulent energy distribution between
scales  (3D energy spectra) of such turbulence. In view of very
sparse experimental data -- besides indirect indications of its
form in He II\cite{VinenNiemela,LSPRE} and
$^3$He-B\cite{Bradley:05} deduced from various decay measurements
or computer simulations\cite{ATN,BrachetNore,TsubotaGP2005} we
recall the only direct measurement of the velocity spectrum based
on pressure fluctuations in He II by Maurer and Tabeling\cite{MT}
-- there is a clear call to tackle this important issue
theoretically. We consider the simplest case of classically
generated turbulence in quantum liquids such as He II or $^3$He--B
that can be thought of as isothermal, homogeneous and isotropic.
We work within a framework of the two-fluid model, building on
ideas first introduced by Volovik\cite{Volovik} and
Vinen\cite{Vinen2005}, by further developing our previous
work\cite{LNV,LSspectra}. We stress that our approach does not
directly apply to quantum turbulence generated in superfluid
helium by the thermal counterflow\cite{VinenOld}, which is
anisotropic, being generated thermally, by the temperature
gradient in the channel.

Above the pressure dependent transition temperature
($T_\lambda\approx 2.17$\,K for $^4$He and $T_c\approx 1$\,mK for
$^3$He) both $^4$He and $^3$He are ordinary viscous fluids that
can be described by the Navier-Stokes equations and their
turbulent flow is fully classical. From hydrodynamical viwepoint
normal $^4$He and $^3$He liquids differ from each other mainly
because of their very different values of kinematic viscosity.
While liquid $^4$He above $T_\lambda$ possesses the lowest
kinematic viscosity, $\nu$, of all known fluids\cite{Russ}, of
order $\nu_{4}\approx2\times10^{-4}$~cm$^2$/s  (fifty times
smaller then that of water at room temperature), liquid $^3$He at
millikelvin temperature is a Fermi liquid\cite{He3}
($\nu_{3}\propto T^{-2}$) with kinematic viscosity exceeding that
of air (which is about $0.15$ cm$^2/$s) comparable with olive oil,
of order $\nu_{3}\approx1$~cm$^2$/s . In principle, both these
liquids may become turbulent.

The plan of the paper is as follows. After this short introductory
Section we introduce the continuous minimal model for two--fluid
turbulence with mutual friction in Section 1 and we use it to
derive the turbulent energy spectra in Section 2. We discuss the
applicability of this continuous model and specify the crucial
role of Kelvin waves in Section 3. We conclude and outline the
future work in Section 4.

 \section{  MINIMAL MODEL  FOR TWO-FLUID
TURBULENCE  WITH MUTUAL FRICTION }

 \subsubsection*{\label{ss:BEM}{\rm 1A.}  Basic equations for the
two-fluid model} In this paper we adopt  the simplest form of the
two fluid model for superfluid $^3$He and $^4$He (see e.g. Eqs.
(2.2) and (2.3) in the Donnelly's textbook\cite{Don})  which
neglects both bulk viscosity and thermal conductivity. These  are
the Euler Eq. for the superfluid velocity $\B u\sb s$ (with zero
viscosity $\nu\sb s=0$) and the Navier-Stokes Eq. for the normal
component $\B u\sb n$ (with the kinematic viscosity $\nu\sb n\=
\nu$):
\BSE\label{NSE} \BEA\label{NSEa} 
&& \rho\sb s \Big[\frac{\p \,\B u\sb s}{\p t}+ (\B u\sb s\* \B
\nabla)\B u\sb s \Big] - \B \nabla p\sb s= -\B F \sb {ns}\,, 
 \\ \label{NSEb}
&& \rho\sb n \Big[\frac{\p \,\B u\sb n}{\p t}+ (\B u\sb n\* \B
\nabla)\B u\sb n  \Big]- \B \nabla p\sb n-\rho\sb n \nu \Delta \B
u_i + \B F \sb {ns}\ . ~~~~~~~\EEA 
Here $\rho\sb n$, $\rho\sb s$   are  the densities of the normal
and superfluid components, $p\sb n$,  $p\sb s$  are
corresponding
 pressures: 
\BEA
  \label{NSEc}
 p\sb n&=&\frac{\rho\sb n}{\rho }[p+\rho\sb s|\B u\sb s-\B u\sb
n|^2]\,,\\
 \label{NSEd}
 p\sb s&=&\frac{\rho\sb s}{\rho }[p-\rho\sb n|\B u\sb s-\B u\sb
n|^2]\,,
  \label{NSEe} \EEA
Here $\rho\= \rho\sb s+\rho\sb n$  and
   $\B F\sb{ns}$ describes the mutual friction:
\begin{equation}
\B F\sb {ns}= -\rho\sb s\{\alpha'({\B u\sb s} - \B u \sb n)\times
\B \omega+ \alpha~\hat{\B \omega\sb s }\times[\B \omega\sb s
\times( \B u \sb s -\B u \sb n) ]\}\  . \label{Fns}
\end{equation}
  $\B \omega\sb s \=\nabla\times \B u\sb s $ is the superfluid
vorticity; $\hat{\B \omega}\sb s \=\B \omega\sb s /\omega\sb s $;
$\alpha'$ and $\alpha$ are dimensionless parameters describing the
mutual friction between superfluid and normal components of the
liquid mediated by quantized vortices which transfer momenta from
the superfluid to the normal subsystem and vice versa. For the
flow with vortices locally aligned with each other these
parameters enter the reactive and dissipative forces acting on a
vortex line as it moves with respect to the normal component.
Here we consider $\alpha'$ and $\alpha$ as phenomenological
parameters, assuming the general case where quantized vortices are
not aligned locally and thus the bare parameters are renormalized.

Following Ref \onlinecite{LNV} we approximate  the ``mutual friction term"
$\B F\sb {ns}$, \REF{Fns},  as follows: 
\BE  \label{NSEf}
  F\sb {ns}=Q \o_0(\B u\sb s -\B
u\sb n) \ . \EE
Here  $Q\simeq \a \rho\sb s$  and $\o_0$ is some characteristic
superfluid vorticity
\BE\label{vort} \o_0\=\sqrt{\<|\B
 \o\sb s|^2\>}\ . \EE
\ESE%
 Approximation \Ref{NSEf} accounts for the fact that the
vorticity in developed  turbulence usually is  dominated by the
smallest eddies in the system  of  scale  $\eta$  with the largest
characteristic wave-vector $k_\eta\sim 1/\eta$. These eddies have
the smallest turnover time $\tau\sb{min}$ that is of the order of
their decorrelation time. On the contrary, the main contribution
to the velocity in the equation for the dissipation of the
$k$-eddies with intermediate wave-vectors $k$,  $k \ll k\sb{max}$,
 is dominated by
the $k'$-eddies with $k'\sim k$. Because the turnover time of
these eddies $\tau_{k'}\gg \tau\sb{min}$, approximation \Ref{NSEf}
adopts self-averaging of \REF{Fns}   on time intervals of interest
($\tau\sb{min} \ll \tau\ll \tau_{k'}$) and thus vorticity can be
considered as almost uncorrelated with the velocity $\B u$
   which is a dynamical variable. Actually,
\REF{NSEf} is the mean field approximation that   neglects   the
fluctuating part of vorticity and   replaces it by   by its mean
value.

\subsubsection*{\label{ss:CC}{\rm 1B.}  Model for the cross-correlation
  $\< \B u\sb s\*\B u\sb n \>  $}
Our first goal in this paper is to formulate the energy balance
equations for the one-dimensional energy spectra   $E_{ii}(k)$ for
the normal and the superfluid subsystems, related with the
simultaneous, same-point correlations $\<\B u_i\* \B u_j\>$ as
follows:
 \BE    \label{def3e}
 \int d k E_{ij}(k)=E_{ij}= 2 \, \<\B u_i\* \B u_j\> \ .
\EE 
\paragraph{a.  Some definitions and known relationships.}

To find the cross-correlation $\<\B u_i\* \B u_j\>$   we need to
recall some definitions and relationships, well known in
statistical physics. The first one is the    Fourier transform:
 \BE \label{def2}
\B u_i(\B k,\o,t) \=  \int \frac{d \B k\,  d \o }{(2\pi)^4}\, \B
u_i(\B r,t) \exp[-i(\B k\* \B r-\o t)]\,, \EE 
and the definition of the two-point, different-time
cross-correlation functions $F_{ij}(\B k,\o)$:
 \BSE\label{def3}
\BEA \label{def3a} && \<\B u_i(\B k,\o)\* \B u_j^*(\B k',\o')\>
\br &\=&  (2\pi)^4 \d(\B k- \B k')\,\d(\o-\o')F_{ij}(\B k, \o)\ .
 \EEA
 Next we need to define simultaneous two-point (cross) correlators,
  $F_{ij}(\B k)$, in $\B k$-representation: 
  \BE \label{def3b}
\<\B u_i(\B k, t)\* \B u_j^*(\B k',t)\> \=  (2\pi)^3 \d(\B k- \B
k')\, F_{ij}(\B k)\ .~~~
 \EE
The frequency integral of $F_{ij}(\B k,\o)$ produces
 $F_{ij}(\B k)$:
 \BE   \label{def3b}
  2\pi\, F_{ij}(\B k) =  \int  d\,\o F_{ij}(\B k,\o)\ .
 \EE
Further  $\B k$-integration gives same-point correlations, which
has a sense of (twice) kinetic energy density per unite mass of
the normal (for $i=j=n$) or the superfluid
 (for $i=j=s$) kinetic energy density per unite mass:
\BE \label{def3c}
(2\pi)^{-3} \int  d^3 k F_{ij}(\B k)= \<\B u_i\* \B u_j\>\= 2
E_{ij}\  . 
\EE

 Our approach is formulated in terms of one-dimensional density of
 $E_{ij}$ in $k$-space,
\BE \label{def3d} 
E_{ij}(k) =   k^2  F_{ij}(k)/ (2\pi)^2\,, 
\EE
 defined such that
  \BE    \label{def3e}
 \int d k E_{ij}(k)=E_{ij}\ .
\EE
 \ESE
Finally, we define a scalar version of the velocity Green's
(response) function: 
\BE\label{defGF} \<\frac{\d u_i (\B k ,\o)} {\d f_j (\B k'
,\o')}\>= (2\pi)^4 \d (\B k-\B k')\,\,\d(\o-\o')\, G_{ij}(\B
k,\o)\ .\EE

 To determine   $\<\B u_i\* \B u_j\>$   we adopt a few more or
less justified assumptions, discussed below.

\paragraph{b. Fluctuation-dissipation approximation for the diagonal
turbulent correlation function.}

In the thermodynamical equilibrium, the different-time (cross)
correlation functions (in the $\o$-representation) are related
with the Greens' function by the mean of the
fluctuation-dissipation theorem, which in hydrodynamics reads: 
\BE\label{FDTe} 
F_{ii}(\B k ,\o)= 2\, T \, \mbox{Im}\, 
[G_{ii}(\B k ,\o)]\,,  \EE 
where $T$ is the temperature of the system. Having in mind that
the thermodynamical equilibrium is the equipartition of the energy
with $T/2$, being the energy per degree of freedom (which in our
notation is $F_{ii}(k)/2$), we can write in equilibrium:
\BE\label{FDT} F_{ii}(\B k ,\o)=  2 F_{ii}(\B k)\, \mbox{Im}  
[G_{ii}(\B k ,\o)]\ . \EE
 Our conjecture  is that this equation is approximately (on a
 semi-qualitative level) valid  also in the flux-equilibrium
 state, i.e., in the fully developed turbulence.

 \paragraph{c. One-pole approximation for the diagonal Greens
functions.} 
Generally speaking, the Greens function in the developed
turbulence is very involved function of $\B k $ and $\o$. The only
facts that we know for sure are:

 1. $ G_{ii}(\B k, \o)$ is analytical in the lower half-plane.

 2. $\displaystyle  \int _{-\infty}^\infty G_{ii}(\B k, \o)\,
d\o=\pi\,, \quad
  \lim_{\o\to\infty}G_{ii}(\B k, \o) = 1/\o\, $

3. At given $\B k$, $ G_{ii}(\B k, \o)$ has a characteristic width
(in $\o$), of about the total damping frequency in the system,
i.e., $ \nu_i k^2+ \gamma_i(k)+ q_i \o_0$.

The simplest analytical from, which satisfies these requirements
is the so-called one-pole approximation 
\BSE\label{approx1} \BE \label{approx1a} G_{ii}(\B k, \o)\approx
\frac 1{ \o -i\left [ \nu_i k^2+ \gamma_i(k)+ q_i \o_0\right]
}\,,\EE 
widely used in theoretical physics in general and in the theory of
turbulence in particular. In \REF{approx1} $ \gamma _i (k)$ is the
eddy-decorrelation frequency, which is the same as the
eddy-turnover frequency. The Kolmogorov 1941 (K41) approximation
for  this object reads:
 \BE \label{approx1b} \gamma _i (k) \approx  C_1 \ve_i (k)^{1/3}
k^{2/3}\,, \EE
\ESE 
where $\ve_i (k)$ is the energy flux in the $i$-subsystem. The
well known effective turbulent viscosity $\nu\Sb T(k)$ is related
with $\gamma (k)$ in the following way: $ \nu\Sb T(k)\= \gamma
(k)/\,k^2 \propto k^{-4/3}$.

 \paragraph{d. Approximation of the Gaussian statistics.}
In the limit $\rho_n\gg \rho _s$  the turbulent velocity $\B u_n$
(or  $\B u_s$ in the opposite limiting case,  $\rho_s\gg \rho _n$)
can be considered as a given one, independent of  $\B u_s$ (or $\B
u_n$, for  $\rho_s\gg \rho _n$). Having in mind that in practice
we are dealing with moderate
 extend of the inertial interval, and that  the statistics of
 turbulence in the energy contained interval is very close to the
 Gaussian one, we can  approximate the statistics of turbulent field
  $\B u_n$ (or $\B u_s$, for  $\rho_s\gg \rho _n$) as
 Gaussian.

\paragraph{e. Here a frog jumps.}
As we suggested, in the limit  $\rho_n\gg \rho \sb s$  we can
consider the cross-velocity term $q\sb s \o_0 \B u_n\= \B f\sb s$
in the RHS of Eq.~\Ref{NSE} for $\B u\sb s$   as a Gaussian random
force $f\sb s$ and compute the cross-correlation (for simplicity
in the scalar version), $\< u\sb s f\sb s \>$,  using so called
Gaussian integration by parts:  
$ \< u\sb s f\sb s^* \>= \<  \d u\sb s/\d f\sb s \>\<f\sb s f\sb
s^*\>$.
 With   $f\sb s= q\sb s \o_0   u_n $  this gives for the velocity
 cross-correlation: 
  \BEA\label{GP} F_{sn}(\B k,\o)&=& q\sb s \o_0
F_{nn}(\B k,\o)G_{ss}(\B k,\o)\ . \EEA  
Using our approximation~\Ref{FDTe} for $F\sb {nn}(k,\o)$,
approximation~\Ref{approx1a} for  $G\sb {nn}(k,\o)$ and $G\sb
{ss}(k,\o)$ relations~\Ref{def3b} and~\Ref{def3c}, one gets after
$\o$-integration: 
\BSE\label{lim5} 
\BE \label{lim5a} E_{sn}(k)  \=  q\sb s \o_0 E_n (k)\Big / \D_k\,,
\quad \mbox{for}  \quad  \rho_n\gg \rho\sb s \ ,
\EE
where $\Delta_k$ is given below by~\Ref{lim1a}. 
Similarly, in the opposite limiting case:
\BE \label{lim5b} E_{sn}(k) \=  q\sb n \o_0 E\sb s (k)\Big /
\D_k\,, \quad \mbox{for}  \quad \rho_n\gg \rho\sb s \ .
\EE
\ESE 
For an arbitrary relation between $\rho\sb n$ and $\rho\sb s$ (including
$\rho\sb s\sim \rho \sb n $)  we suggest  the interpolation
formula \BSE\label{lim1}
 \BEA E_{sn}(k)  &=& \o_0[q\sb n  E\sb s(k)+ q\sb s E\sb n(k)]
\Big / \D _k\,,\label{lim1b} \\ \label{lim1a} \Delta  _k &\=&
\nu\, k^2 + \gamma\sb n(k)+ \gamma\sb s(k)+ (q\sb s+q\sb
n)\o_0\,,
\EEA\ESE
 which obeys all needed limiting cases. Moreover, in the limiting
 case $\o_0 q_i\gg \gamma _i$ it turns into a simpler form
\BSE\label{lim2}
 \BE\label{lim2a}
E_{sn}(k)  = [\rho\sb s  E\sb s(k)+ \rho\sb n E\sb n(k)] / \rho \
.\EE 
This equation has a physically motivated solution
\BE \label{lim2b} E_{sn}(k)=E_{s }(k)=E_{ n}(k) \,, \EE
that gives $\< \B u\sb n(\B k, t)[\B u\sb n(\B k, t)-\B u\sb s(\B
k, t) ]\> =0$
 and thus requires 
\BE \label{lim2d}
 \B u\sb n(\B r, t)=\B u\sb s(\B r, t)\,,\EE
\ESE i.e., a fully coherent motion of the superfluid and the
normal fluid velocities. It means that our
interpolation~\Ref{lim1} is physically justified and actually
works better than one would expect, having in mind our rather
crude approximations.

\subsubsection*{\label{ss:bal} {\rm 1C.} The energy balance equations}
 The energy balance equations can  be derived  in a standard manner
 from the equations of motion,
 Eqs.~\Ref{NSE} -- see e.g. Ref.~\onlinecite{LNV}. 
 The only new factor in this derivation is the cross-velocity
 correlations, for which we adopt Eqs.~\Ref{lim1}. This gives:
\BSE \label{EB}
\BEA \label{EBa}
&& \frac{\p E\sb n(k)}{\p t}+ \frac{\p \ve\sb n (k)}{\p k}+ \nu
k^2\br 
&=& q\sb n \o_0 \Big\{\frac{\o_0}{\D_k}\big[q\sb sE\sb n(k)+q\sb
nE\sb s(k)\big]- E\sb n(k) \Big\}
\\
\label{EBb} && \frac{\p E\sb s(k)}{\p t}+ \frac{\p \ve\sb s
(k)}{\p k} \br 
&=& q\sb s \o_0 \Big\{\frac{\o_0}{\D_k}\big[q\sb sE\sb n(k)+q\sb
nE\sb s(k)\big]- E\sb s(k)   \Big\} \,, \EEA\ESE 
with $\D_k$, given by \REF{lim1a}. The simplest way to model
$\ve_i(k)$, suggested in Ref.~\onlinecite{Kov},  is to relate
$E_i(k)$ and $\ve_i(k)$ in \REF{EB} in the spirit of the K41
dimensional reasoning:
\begin{equation}\label{V5} 
E_i(k)=C \ve_i(k)^{2/3}  k^{-5/3}\ .
\end{equation}
Here  $C\simeq 1$ is the Kolmogorov dimensionless constant. In the
absence of dissipation,  \REF{V5} immediately produces the
stationary solution $\ve_k=\ve$ with constant energy flux $\ve$ in
the inertial interval of scales. Then  \REF{V5} turns into  the
Kolmogorov-Obukhov $5/3$--law for $E_k$:
\begin{equation}\label{V6} 
E_k=C \ve^{2/3} k^{-5/3}\ .
\end{equation}

\section{\label{s:spectra}  TURBULENT ENERGY SPECTRA IN THE
MINIMAL MODEL}

\subsubsection*{\label{ss:small-n}{\rm 2A.} Small normal density}

Let us first consider the case
\BE\label{as1} \rho\sb s\gg \rho\sb n\,, \ \mbox{and thus:}\ \ \
q\sb s\ll q\sb n\ . 
\EE
Clearly, in this case the massive superfluid component does not
feel the tiny  superfluid one and thus in the zeroth-order
approximation [with respect of $(\rho\sb n/\rho\sb s)\ll 1$] 
\BE\label{res1} 
E\sb s(k)=C \ve\sb s ^{2/3}k^{-5/3}\,, \ \ \mbox{K41 spectrum}\
.
\EE 
Indeed, in the limit $\rho\sb s\to 0$ one has zero in the RHS of
\REF{EBb}. Then in the stationary case $\p \ve\sb s/ \p k=0$,
i.e., $\ve\sb s$ becomes $k$-independent, and from \REF{V5} one
immediately gets \REF{res1}.

Much more interesting question is about a spectrum of the normal
component of the small density, which is essentially affected by
the massive superfluid component.  In the limit~\Ref{as1} and for
the stationary case Eqs. \Ref{EBa} and \Ref{lim1a} for $E\sb n
(k)$ take the form
\BSE \label{BEn}\BEA\label{BEna} && \hskip -1cm \frac{\p \ve\sb n
(k)}{\p k} +
 \nu k^2 E\sb n(k) = q\sb n \o_0 \Big[\frac{\o_0
q\sb nE\sb s(k) }{\D_k} - E\sb n(k) \Big],~~~   \\
 && \Delta  _k  \=  \nu\, k^2 +
\gamma\sb n(k)+ \gamma\sb s(k)+  q\sb n \o_0\,, \label{BEnb}
\EEA\ESE 
that will be analyzed for the two limiting cases.
\paragraph{\label{sss:smallk} a.  Small $k  \Rightarrow$
 full coupling}
For small $k$, $\gamma\sb s$ (being $\propto k^{2/3}$) is small
with respect of the ($k$-independent) $\o_0 q \sb n$. If so,
$\D_k\to q\sb n \o_0$ and 
\BSE\label{res2}
 \BE\label{res2a}  \p \ve\sb n (k)/\p k =q\sb n[
E\sb n(k)-E\sb s(k)]\,, 
\EE
with the K41 solution
\BE\label{res2b} 
E\sb s(k)=E\sb n(k)=C \ve^{2/3}k^{-5/3}\,, \ \ \ve\sb s=\ve\sb
n\=\ve\,,\EE\ESE
 having full  coupling of the velocities, $\B u\sb s=\B u\sb n$.

 \paragraph{\label{sss:smallk} b.  Large  $k  \Rightarrow$
decoupling with K41 regime} 
K41 viscous micro-scale, $k_\eta\= 1/\eta$,
 defined in the standard manner:
 \BE\label{eta} 
\nu k_\eta^2 = \gamma\sb n(k_\eta)\,, \ \ \Rightarrow k_\eta
\simeq
 \ve\sb n^{1/4}/\nu^{3/4}\ . 
 \EE
In absence of superfluid component, the
spectrum of developed turbulence of the normal fluid would
decay exponentially for $k\gg k_\eta$. We demonstrate that in the
two-fluid system  this does not occur due to an additional energy
flux from the superfluid component to the normal  one.  Let
$k>k_\eta$. Then one simplifies Eqs.~\Ref{BEn} to an algebraic
form with solution 
\BSE\label{sol1} \BE\label{sol1a} 
E\sb n(k)=E\sb s(k)\frac{q\sb n^2 \o_0^2}{(\nu\, k^2+ q\sb n
\o_0)^2}\ . \EE One finds here a new scale: 
\BE\label{sol1b} \nu\, k_*^2= q\sb n \o_0\ .
\EE
\ESE

\BSE  \label{lim3}
  In the case of essential mutual friction, when  $ k_*\gg k_\eta$,
(i.e.   $ q\sb n \gg \nu k_\eta^2$),  in the subinterval
\BE\label{lim3b} k_*\gg k \gg k_\eta\,, \EE
 a  spectrum of normal  component (in the former viscous interval!)
deviates from the K41 just slightly:
\BEA \label{lim3c} 
E\sb s(k)-E\sb n(k)&\approx & E\sb s(k) \frac {\nu^2 k^4}{q\sb
n^2\o_0^2}\ll 1 \
 . \EEA 
It means that in the subinterval~\Ref{lim3b} the mutual friction
is still strong enough to keep the  normal velocity almost coupled
to the superfluid one: 
\BE   \label{lim3d}  |\B u\sb s- \B u\sb n| \approx  u\sb s
 \frac {\nu^2 k^4}{2\,q\sb n^2\o_0^2}\ll u\sb s \ .
\EE \ESE

 However, for
 \BSE \label{lim4}
 \BE \label{lim4a}
 k> k_*
 \EE
 the strong coupling~\Ref{lim3d}  disappears, as it follows from \REF{sol1a}.
 In this region, according to \REF{sol1a}: 
\BE \label{lim4b} 
 E\sb s\gg E\sb n \approx E\sb s \frac {  q\sb
n^2\o_0^2} {\nu^2 k^4}\propto k^{-4 -5/3}\,, \EE
 \ESE
i.e. $E\sb n \ll E\sb s$. However,  in contrast with standard K41
spectrum, one has here a power-law rather than an exponential
decay.
\begin{figure}
 \includegraphics[width=6cm]{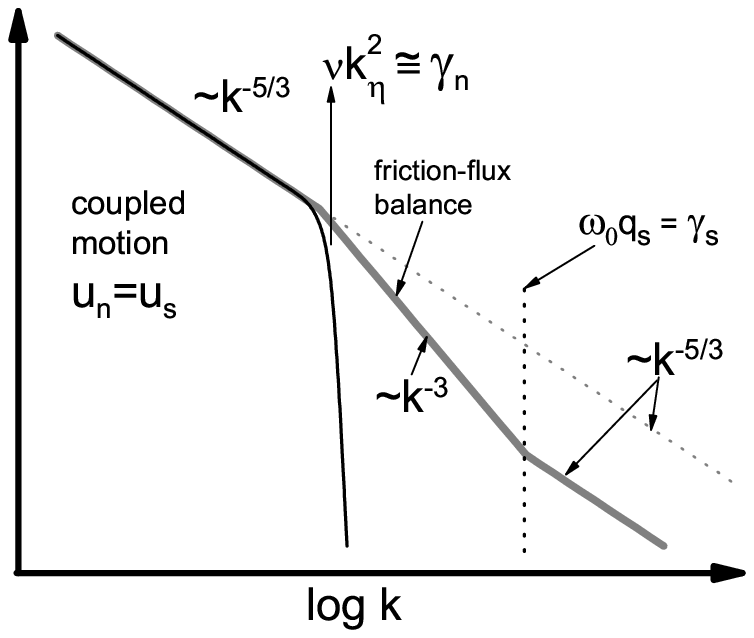}
  \includegraphics[width=6cm]{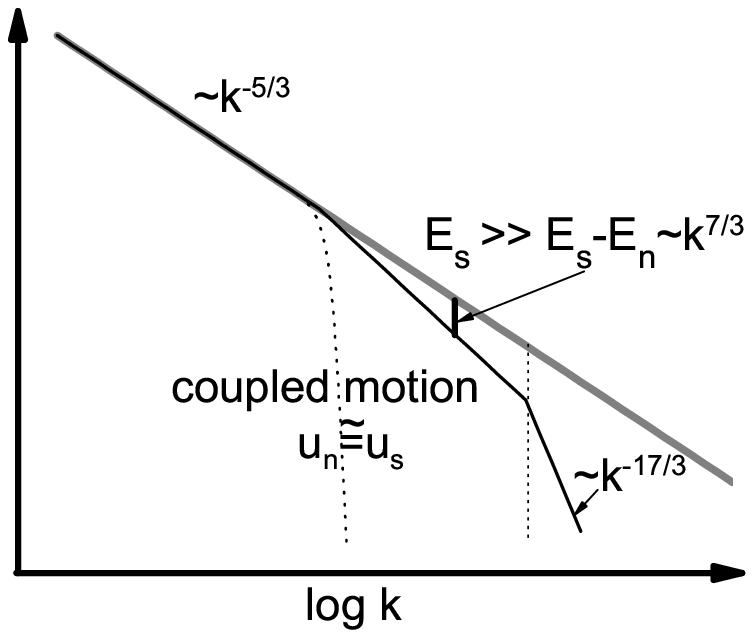}
  \includegraphics[width=6cm]{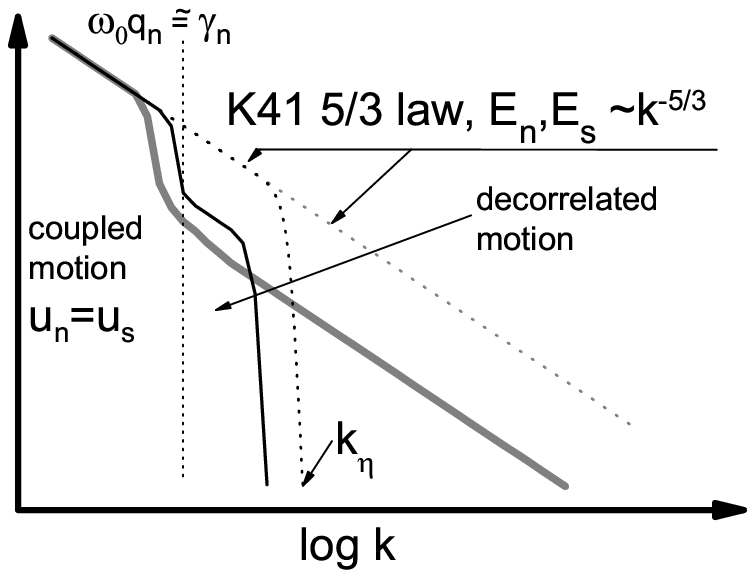}
 \includegraphics[width=7cm]{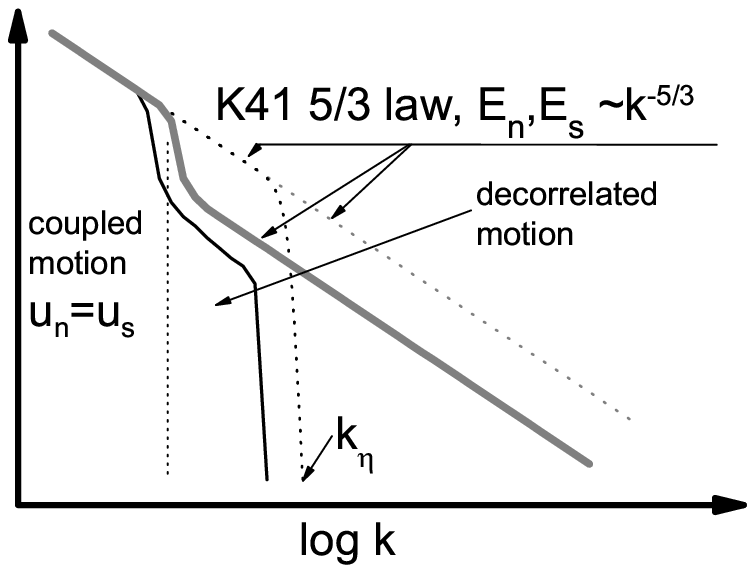}
  \caption{The asymptotic forms of the coupled 3D energy spectra (schematic log-log plots of energy spectral density versus wave number)
  in the normal fluid (solid black lines) and in the superfluid (solid grey lines)
  resulting from the continuous two-fluid model. The  dotted lines indicate the form of original conventional
  uncoupled K41
  spectra. The top (bottom) figures correspond
  to limiting cases of large (low) mutual friction, for large (left) and small (right) normal fluid density.
  For further details, see text.}\label{f:2}
\end{figure}

 \subsubsection*{\label{ss:small-n}{\rm 2B.} Small superfluid  density}

Consider now the opposite  limiting case to \REF{as1}
\BE\label{as9} \rho\sb s\ll \rho\sb n\,, \ \mbox{and thus:}\ \ \
q\sb s\gg q\sb n\ . 
\EE
Clearly,  then (in the zeroth-order approximation) in the inertial
interval of scales, $k<k_\eta$, the normal component obeys the
standard K41 spectrum
\BE\label{res10} 
E\sb n(k)=C \ve\sb n ^{2/3}k^{-5/3}\,,
\EE 
in the inertial interval of scales, $k<k_\eta$. The spectrum of
the superfluid component is considered below in various limiting
cases.

\paragraph{\label{sss:1}a.  Small mutual friction}
Consider first the case when, in addition to
inequalities~\Ref{as1}, one has
\BSE\label{kdag} \BE \label{kdag-a} 
k_\dag \ll k_\eta\,,
 \EE
where the new characteristic scale $k_\dag$ is defined by
 \BE \label{kdag-b} 
 q\sb s\o_0= \gamma\sb s(k_\dag)\ . \EE \ESE For $k<k_\dag$ the
 mutual friction dominate over the nonlinear interaction. We already know
 (see \REF{lim2d}) that in this case  $\B u\sb n=\B u\sb
 s$, i.e., there is a complete coupling of normal and superfluid
 velocities, both spectra obeying the $5/3$-law~\Ref{res2b}.

 For $k>k_\dag$ the nonlinear interaction dominates over the mutual
 friction:
 \BE \label{res3}
 \< \B u\sb s\* \B u\sb n\> \ll \< |\B u\sb s|^2\>\ .
 \EE 
Nevertheless, both spectra obey $5/3$-law with $\ve\sb s\simeq \ve
\sb n$. Normal spectrum ends at $k\simeq k_\eta$. Superfluid
component does not feel this cutoff and continues further until the
limit of the classical description. Beyond this limit, the energy
cascade is taken over by reconnections and Kelvin waves (see below).

\paragraph{\label{sss:2}b.  Large mutual friction,
} In some sense, richer physics corresponds to the case with
larger mutual friction, when 
\BE\label{as2}
 k_\dag \gg k_\eta \ .
\EE 
Then one has a coupled turbulent motion of the normal and
superfluid components (with $ \B u\sb n(\B r, t)=\B u\sb s(\B r,
t)$) until the viscous cutoff~\Ref{eta}.

For \BE k_\eta < k < k_\dag\EE one has a case of strong mutual
friction with the normal component  at rest. This case has been
considered in details in our Ref.~\onlinecite{LNV} (and agrees
with numerical result of Vinen\cite{Vinen2005}), giving the
$-3$-spectrum
\BE\label{LNV} E\sb s(k)\simeq q\sb s^2 \o_0^2  k^{-3}\,, \EE 
that originates from the balance of the nonlinear flux and the
friction terms in \REF{res4} below.

Indeed, in this case one puts in \REF{EBb} $E\sb n=0$, $\D_k\simeq
\o_0 q\sb s$, and the RHS of \REF{EBb} can be approximated as
$$\mbox{RHS}\approx \o_0 q\sb s E\sb s \Big(\frac{q\sb n}{q\sb s}
-1\Big)\approx - \o_0 q\sb s E\sb s\ .
$$
In that case, instead of \REF{EBb}, one arrives at 
 \BE\label{res4}
 q\sb s \o_0 E\sb s (k) = \frac{\p \ve\sb s(k)}{\p k}= C^{-3/2}\,
 \frac{\p}{\p k}\big[k^{5/2}E\sb s^{3/2} (k)\big]\,,
\EE 
which has the solution~\Ref{LNV}.

For $ k > k_\dag$ the mutual friction is NOT important and the
superfluid component recovers the $5/3$-law, but with smaller
energy
flux.\\~\\

\section{ROLE OF KELVIN WAVES}

The description we presented above ignores the fact that
turbulence of the superfluid component consist of discrete tangled
vortices with quantised circulation $\kappa$. This description is
valid for the scales greater that the mean distance $\ell$
separating the quantised vortices. For example, in $^4$He above
1.5~K the normal fluid Kolmogorov scale appears to be of the same
order as $\ell$ and, therefore, in this case our model is not
applicable for predicting the superfluid spectrum below the
Kolmogorov scale\cite{VinenNiemela}. Another example is turbulence
at $T<1K$ when the normal component is extremely weak and behaves
like a Knudsen gas rather than a fluid. In this case, most of
energy reaches $\ell$ along the cascade without dissipation and
one should ask what happens to this energy below this scale. At
scale $\ell$, an essential role in turbulence evolution play
vortex reconnections during which part of the energy is lost to
phonon emission, and the rest of the energy is transferred to
Kelvin waves, see e.g. paper of W.F. Vinen in this issue. The
reconnections produce sharp cusps which quickly transform into a
superposition of Kelvin waves whose nonlinear interaction leads to
further turbulent cascades through scales. Note that these sharp
cusps correspond to a broad distribution in wavenumber space and,
therefore, both the direct energy cascade and the inverse cascade
of waveaction can be important in subsequent
evolution\cite{nazarenko}. To describe the statistical nonlinear
Kelvin waves one can use the weak turbulence approach, which
results in a six-wave kinetic equation for the energy
spectrum\cite{kozik}. A differential equation model for Kelvin
wave turbulence (kelvulence) which preserved the essential
scalings and solutions of the original integral kinetic equation
was derived in Ref~\onlinecite{nazarenko}:
\begin{equation}
\dot n = {C  \over \kappa^{10}} \omega^{1/2} {\partial^2 \over
\partial \omega^2} \left(
 n^6 \omega^{21/2} {\partial^2 \over \partial \omega^2} {1 \over n}
\right), \label{forth}
\end{equation}
where $n=E_s L/\omega$ is the waveaction spectrum\footnote{Factor
$L$ appears when one calculates energy per unit volume in terms of
the energy per unit vortex length}, $L$ is the vortex length per
unit volume -- the vortex line density,
 $\kappa$ is the circulation quantum, $C$ is a dimensionless constant
and $\omega = \omega (k) \cong {\kappa  k^2/ (4 \pi)}$ is the
Kelvin wave frequency\footnote{Here, we ignore logarithmic
factors}. This equation preserves the energy (per unit length of
the vortex)
\begin{equation}
E = {1 \over 2 \sqrt{\kappa}} \int \omega^{1/2} n \, d\omega
\end{equation}
and the waveaction  (per unit  length of the vortex)
\begin{equation}
N = {1 \over 2 \sqrt{\kappa}}  \int \omega^{-1/2} n \, d\omega\ .
\end{equation}
Equation (\ref{forth}) has both the direct cascade solution $n
\sim k^{-17/5}$  and  the inverse cascade solution
 $n \sim k^{-3}$.
It also has the family of thermodynamic Rayleigh-Jeans solutions,
\begin{equation}
n = { T \over  \omega + \mu}\ . \label{term}
\end{equation}
where $T$ and $\mu$ are constants having a meaning of temperature
and the chemical potential, respectively.

The direct cascade of energy in kelvulence is eventually
dissipated at high wavenumbers either via phonon radiation or via
friction with the normal component. The phonon radiation is the
dominant dissipation mechanism near absolute zero whereas the
mutual friction becomes more important at higher temperatures,
e.g., at $T>0.4$~K in $^4$He, as estimated in Ref
\onlinecite{VinenNiemela}. Using a dimensional argument and an
assumption that the radiation is quadrupolar, we have the
following expression for the sound dissipation
term\cite{nazarenko},
\begin{equation}
(\dot n)_{rad} = - { \omega^{9/2} n^2 \over  \kappa^{1/2} c_s^4},
\label{soundrad}
\end{equation}
where $c_s$ is the speed of sound. This kind of sound radiation
terminates the energy cascade at a finite frequency $\omega_{\rm
rad} \sim (\epsilon^3 c_s^{20} / \kappa^{16})^{1/13}$, where
$\epsilon$ is the total energy injection rate per unit vortex
length\cite{nazarenko}.

Now let us introduce the effect of mutual friction. It was argued in
Ref~\onlinecite{VinenNiemela} that the characteristic time of the
mutual friction dissipation at low temperatures, when the normal
fluid is at rest, is $\tau_{\rm mf} \sim 1/(\alpha \kappa k^2)$
where $\alpha$ is the (temperature dependent\cite{Russ,Bevan})
mutual friction coefficient. Thus, we can postulate the following
dissipation term in our differential model,
\begin{equation}
(\dot n)_{\rm mf} = - {\alpha \kappa }
 k^2 n. \label{mf}
\end{equation}
Interestingly, this term looks like a viscous dissipation with an
effective viscosity coefficient $\alpha\kappa$. Its temperature
dependence qualitatively agrees (within a numerical factor of
order unity) with measurements of the effective kinematic
viscosity in $^4$He extracted from the towed grid experiments in
the range between about 1.2~K to 1.7~K\cite{stalp}. For higher
temperatures, coefficient $\alpha$ has to be replaced with a more
complicated function of both friction parameters $\alpha$ and
$\alpha'$~\cite{VinenPrivate}. For a finite counterflow velocity
$V_{\rm ns}$, one should replace (\ref{mf}) with\cite{BarTsub}
\begin{equation}
(\dot n)_{\rm mf} = \alpha \left[k V_{\rm ns} - \kappa k^2
{\log(1/ka) \over 4 \pi} \right] n\, , \label{mf1}
\end{equation}
which describes Glaberson instability of Kelvin waves when their
phase velocity is less than $V_{\rm ns}$. Finally, we are leaving
for future consideration an interesting case of turbulent Glaberson
amplification when $V_{\rm ns}$ is random due to turbulence in the
normal fluid. Note that the friction dissipation in (\ref{mf1})
differs from (\ref{mf}) by factor ${\log(1/ka)\big / 4 \pi} $ which,
following Ref~\onlinecite{VinenNiemela}, we will assume to be close
to unity.

To examine the effect of mutual friction on the energy cascade,
let us, following Ref.~\onlinecite{nazarenko}, introduce a reduced
second order model which ignores the waveaction conservation and
the inverse cascade. With the friction term (\ref{mf}) included,
we have
\begin{equation}
\dot n = - {C_1  \over \kappa^{10}\sqrt \omega}  {\partial \over
\partial \omega} \left(
 n^5 \omega^{17/2}
\right) - {4 \pi \alpha } \omega n. \label{first}
\end{equation}
where $C_1>0$ is a  dimensionless constant. The general stationary
solution of equation (\ref{first}) is
\begin{equation}
 n =  \left[ \left( {2 \epsilon \over C_1} \right)^{4/5} \kappa^{42/5} -
 {4 \pi \alpha \kappa^{10} \over C_1} \, \omega^{4/5}
 \right]^{1/4}\, \frac 1 {\omega^{17/10}}\  .
\label{mfspec}
\end{equation}
We see that at low frequencies this expression coincides with the
non-dissipative energy cascade spectrum, $n \sim \omega^{-17/10}$,
and we also see a sharp cut-off at
\begin{equation}
\omega_{\rm mf} = {2 C_1^{1/4} \epsilon \over (4 \pi \alpha)^{5/4}
\kappa^2}\ . \label{mfcut}
\end{equation}
Comparing $\omega_{rad}$ and $\omega_{\rm mf}$ one can see that the
frictional dissipation becomes more important than the phonon
radiation if
\begin{equation}
\epsilon < (4 \pi \alpha)^{13/8} c_s^2 \kappa\  . \label{border}
\end{equation}
This expression differs from expression (87) of
Ref.~\onlinecite{VinenNiemela} (assuming their relation (74)
between $\epsilon$ and $l$, they obtained crossover between the two
regimes at $\alpha\simeq 7\times 10^{-8}$ which corresponds to the
temperature about 0.4~K). We attribute this difference to the fact
that Ref.~\onlinecite{VinenNiemela} assumed the sound radiation to
be dipolar rather than quadrupolar as in
Ref.~\onlinecite{nazarenko}.

Here we present our estimate of the He II temperature when
condition (\ref{border}) indicates a crossover between phonon
radiation and mutual friction dissipation: injecting a power at
the level of 1~W into 1 liter of liquid of density about
145~kg/m$^3$ (i.e., energy decay rate $\varepsilon\approx
7$~m$^2$/s$^3$) results to a typical vortex line density
$10^{10}-10^{11}$~m$^{-2}$, giving $\epsilon\simeq 10^{-10}$~
m$^4$/s$^3$. The condition (\ref{border}) thus requires $4 \pi
\alpha$ of order $10^{-5}$ (i.e., about an order of magnitude
higher than the estimate of Ref.~\onlinecite{VinenNiemela}), which
roughly corresponds\cite{BDV} to temperature of about 0.5~K .

\section{ CONCLUSIONS}

In this paper, we presented a minimal model for turbulence of the
coupled superfluid and normal components in superfluid helium. The
model comprises a system of nonlinear partial differential equations
for the energy spectra and its origin in the case of classical
fluids can be traced back to Kovasznay 1947 paper\cite{Kov}. The
basic idea of such models is that the nonlinear terms, being of the
simplest possible form, should preserve the original turbulence
scalings and, in particular, predict correctly the Kolmogorov
cascade. Having the Kolmogorov scalings built into the model and
adding additional physical interactions, such as  mutual friction
and viscosity, one then obtains new nontrivial physical regimes
charactrerised by non-Kolmogorov spectra. For superfluids, the first
application of such model was done in Ref. \onlinecite{LNV} in the
limiting case of the normal fluid at rest, and the new $-3$ spectrum
was predicted.

In the present paper, we generalised this model to the case where
both the normal and the superfluid components may be turbulent. The
crucial theoretical step here is our estimation of the
cross-correlation function between normal and superfluid velocities
which determines their joint dynamics. This cross-correlations leads
to appearance   of a rich variety of interesting new regimes. In
particular, we found for large superfluid to normal density ratio a
regime in which the normal component is ``dragged'' by the
superfluid component via the mutual friction with resulting
Kolmogorov spectrum extending  far below the Kolmogorov dissipation
scale. Our model also confirms the picture suggested in
Ref.~\onlinecite{LSspectra} of a ``knee'' spectrum where the
dissipative $-3$ scaling at medium wavenumbers exists in between of
$-5/3$ Kolmogorov ranges at low and large $k$'s. Our theory bridges
classical turbulence with quantum turbulence and in a quantitative
manner, it  points out similarities and differences between the two,
and we expect it to be useful and efficient for numerical
simulations of more complicated experimental cases.

We also discuss the case when our continuous two-fluid description
breaks down at scales below the mean distance between the
quantised vortex filaments. At these scales, the turbulent
cascades are believed to be carried through by random nonlinearly
interacting Kelvin waves (kelvulence). A differential model for
kelvulence, including the phonon radiation effect, was proposed in
Ref.~\onlinecite{nazarenko}. In the present paper we extend this
model to include the mutual friction effect and we obtain an
analytical solution where the energy cascade is arrested at a
finite wavenumber. Comparing this friction cut-off with the
previously obtained radiation cutoff\cite{nazarenko} we obtained
an estimate for the crossover between the radiation and the
friction dissipation mechanisms.

We leave the exact details of applicability to He II and to
$^3$He-B for future work, as well as an extension of our approach
to (anisotropic) counterflow turbulence. Let us point out that the
likely candidate of experimental verification of our theory might
be $^4$He-$^3$He superfluid mixtures where, due to presence of
$^3$He quasiparticles, mutual friction at low temperature is still
expected to be significantly higher than in pure He II.

\section{ACKNOWLEDGEMENTS}
Discussions with many colleagues, especially with  C.F. Barenghi,
P.V.E. McClintock, J.J. Niemela, K.R. Sreenivasan, M. Tsubota,
W.F. Vinen and G.E. Volovik are warmly acknowledged. This research
is supported by the research plan MS 0021620835 financed by the
Ministry of Education of the Czech Republic, by the GA\v{C}R under
202/05/0218, by the ULTI-4 and the US-Israel Binational Science
Foundation.

\end{document}